\documentclass{article}
\usepackage{spconf,amsmath,graphicx}
\usepackage{multirow}
\usepackage{booktabs}
\usepackage{amssymb}
\usepackage{bm}

\usepackage{enumitem}
\setlist{nosep, leftmargin=14pt}

\usepackage{mwe} 

\title{Rapid Brain Meninges Surface Reconstruction \\ with Layer Topology Guarantee}
\address{
$^{1}$ Department of Biomedical Engineering, Johns Hopkins School of Medicine, USA\\
$^{2}$ Department of Biomedical Engineering, Yale University, USA\\
$^{3}$ Department of Electrical and Computer Engineering, Johns Hopkins University, USA\\
$^{4}$ Laboratory of Behavioral Neuroscience, National Institute on Aging,\\ National Institutes of Health, USA\\
$^{5}$ Department of Neurology, Johns~Hopkins~School~of~Medicine, USA}
\begin{document}
%
\maketitle
\begin{abstract}
The meninges, located between the skull and brain, are composed of three membrane layers: the pia, the arachnoid, and the dura. Reconstruction of these layers can aid in studying volume differences between patients with neurodegenerative diseases and normal aging subjects. In this work, we use convolutional neural networks~(CNNs) to reconstruct surfaces representing meningeal layer boundaries from magnetic resonance~(MR) images. We first use the CNNs to predict the signed distance functions~(SDFs) representing these surfaces while preserving their anatomical ordering. The marching cubes algorithm is then used to generate continuous surface representations; both the subarachnoid space~(SAS) and the intracranial volume~(ICV) are computed from these surfaces. The proposed method is compared to a state-of-the-art deformable model-based reconstruction method, and we show that our method can reconstruct smoother and more accurate surfaces using less computation time. Finally, we conduct experiments with volumetric analysis on both subjects with multiple sclerosis and healthy controls. For healthy and MS subjects, ICVs and SAS volumes are found to be significantly correlated to sex~(p$<$0.01) and age~(p$\leq$0.03) changes, respectively.
\end{abstract}
\section{Introduction}
\label{sec:intro}

Neuroimage studies focus solely on the brain and ignore regions outside the brain. However, the meninges---located between the brain and the inner table of the skull---play an important role in many pathological conditions~\cite{brain_meninges_anatomy_2021}. It comprises three thin membranes: the pia, arachnoid, and dura. These distinctive layers help to define clinically interesting volumetric measurements for brain studies. The area between the arachnoid and pia, called the subarachnoid space~(SAS), is filled largely by cerebrospinal fluid~(CSF). Although few imaging studies have been conducted on the SAS, it is essential for circulating the CSF, and estimates of its size could help our understanding of brain atrophy~\cite{brain_meninges_anatomy_2021}. A prior study found that SAS volume increases with normal aging~\cite{Duan_SPIE_2022}; however, SAS volumes in patients with certain neurodegenerative diseases could change differently and such changes have not been thoroughly investigated in the literature.
For the dura layer, The intracranial volume~(ICV) can be calculated as the volume enclosed by the dura mater and is an important normalization measure in many neuroimaging studies~\cite{Han-2020-longitudinal}.

The study of SAS and ICV changes in neurodegenerative diseases relies on accurate surface reconstruction and analysis using magnetic resonance images~(MRIs). MR imaging has been widely used in neuroimaging studies due to its soft tissue contrast. It is also common to use multiple MR contrasts since T1w images are typically used to achieve a better contrast between gray matter~(GM) and white matter~(WM) tissue, while T2w images are good at visualizing fluid-tissue contrast~\cite{ZUO_CALAMITI_2021}. The pia lies between the GM and CSF, while the arachnoid lies between the CSF and dura. Both of these are extremely thin membranes and can be accurately represented by a single surface. In contrast, the dura may be $0.5$--$2.0$~mm thick and should be represented by its bounding surfaces. Accordingly, we represent its inner surface by the arachnoid and its outer surface---which lies between the dura and bone---by the epidural surface. We aim to reconstruct arachnoid and dura surfaces with correct anatomical ordering from conventional multi-contrast MRI. 

Reconstructing the meninges from conventional MRI is a challenging task.
As we have noted, the membranes themselves are thin and the meningeal thickness varies with age, sex, and disease~\cite{meninges_vary_thickness_1, meninges_thickness_2}. Previous work used the nested topology preserving geometric deformable model~(NTGDM) to enforce anatomical ordering in combination with a convolutional neural network~(CNN) to achieve sub-voxel accuracy to reconstruct the arachnoid and dura surfaces~\cite{Duan_SPIE_2022}. However, NTGDM requires a multiple-step optimization procedure, which is computationally expensive and may not be feasible in large-scale studies.

Our goal is to generate smooth arachnoid and dura surfaces more efficiently while maintaining the accuracy of the surface reconstruction. In this paper, we propose a new approach that produces nested topology-preserving surfaces from deep networks by applying ReLU activation to the output channels of a CNN~\cite{HE_2021_OCT} to ensure non-negative difference maps between adjacent layers. The network uses paired T1-weighted~(T1w) and T2-weighted~(T2w) MRIs to estimate the arachnoid and dura layers. Our single-step feed-forward CNN generates signed distance functions~(SDFs)~\cite{malladi1995tpami}, while the marching cubes algorithm was used to reconstruct surfaces from the SDFs. The pia SDF is used to facilitate the nested topology via SLANT-CRUISE~\cite{Huo-2016-MACRUISE,Huo-2019-slant}. By conducting experiments on a normal aging cohort and a multiple sclerosis~(MS) cohort, we present a pilot study of volumetric analyses in changes of ICV and SAS volumes for both healthy subjects and patients with neurodegenerative diseases.

\begin{figure*}[!t]
\centering
  \includegraphics[width=0.98\linewidth, trim={0 8cm 0 10cm}]{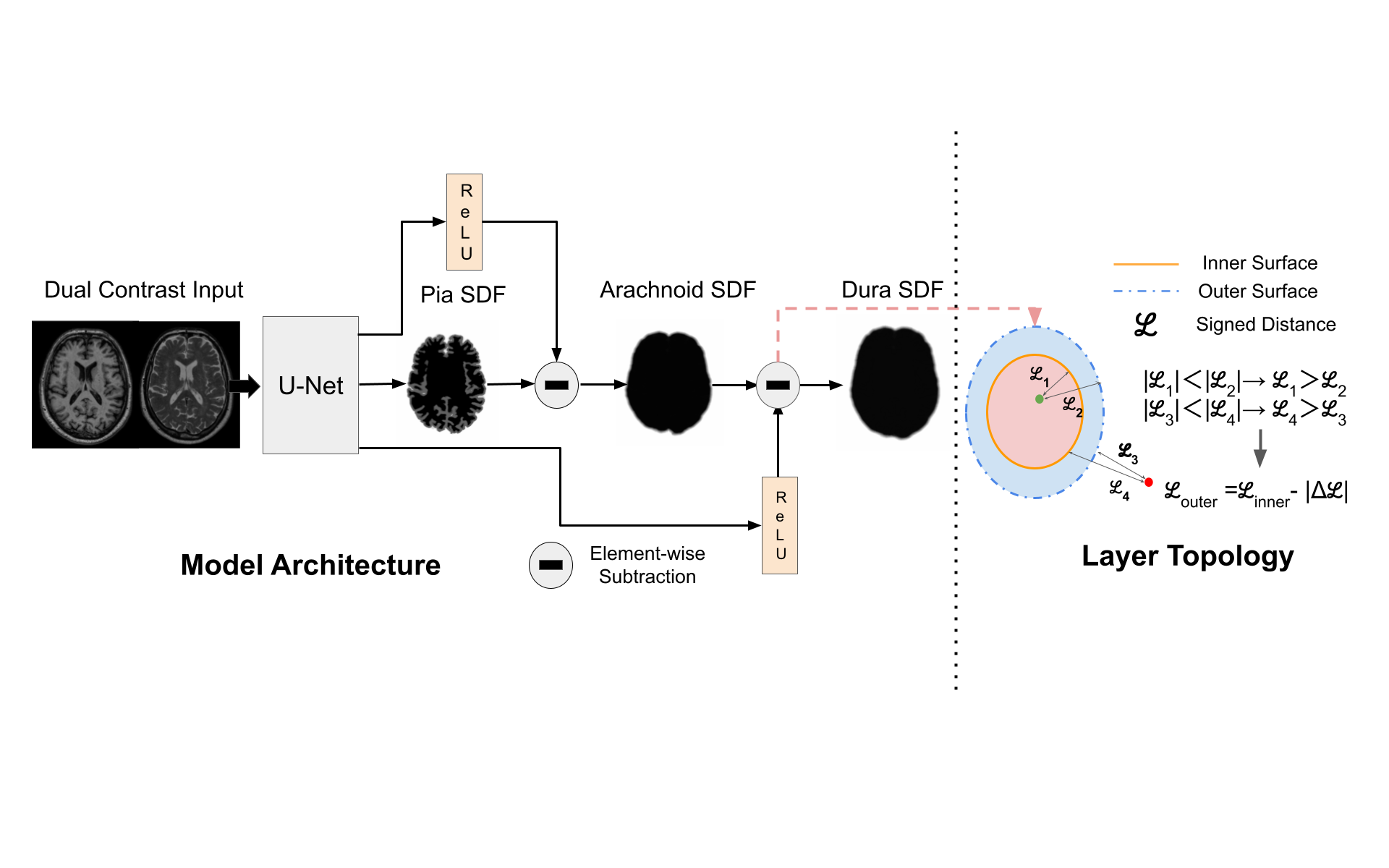}
  \caption{Architecture of our proposed surface reconstruction model with its layer topology guarantee.}
  \label{fig:Archetecture}
\end{figure*}

\section{Method}
To accurately reconstruct the meninges layer surfaces, we use a 2D U-Net that takes pairs of T1w and T2w axial slices and outputs three SDFs, as shown in Fig.~\ref{fig:Archetecture}. To preserve the layer topology, distances between adjacent meningeal layers are guaranteed to be non-negative by using ReLU activation. More details are introduced in the remainder of this section.

\noindent\textbf{Nested Layer Topology}~~~~Our network directly outputs an SDF for each individual meningeal surface. An SDF, denoted as $\phi(x)$, is defined as the shortest distance between point $x$ and the surface when $x$ is outside the surface and the negative of that distance when inside the surface. Since the pia surface is inside the arachnoid surface, and the arachnoid surface is inside the epidural surface, we demonstrate in the right-hand side of Fig.~\ref{fig:Archetecture} that for any point x, $\phi_{pia}(x) \ge \phi_{ara}(x)$ and $\phi_{ara}(x) \ge \phi_{dura}(x)$ where $\phi_{pia}$, $\phi_{ara}$, and $\phi_{dura}$ are the SDFs of pia, arachnoid, and dura, respectively. Therefore, we could guarantee the nested layer topology by enforcing: 
\begin{equation}
    \label{eq:sdf_diff}
    \Hat{\phi}_{\text{pia}} - \Hat{\phi}_{\text{ara}} \ge 0 \quad \text{and} \quad 
    \Hat{\phi}_{\text{ara}} - \Hat{\phi}_{\text{dura}} \ge 0,
%
%
\end{equation} 
where $\Hat{\phi}$ is the estimated SDF from our proposed model. 

To preserve the topology between the surfaces, the constraints in Eq.~\ref{eq:sdf_diff} must be met. Thus, the difference between the inner surface SDF and the outer surface SDF $\Delta\phi_{\text{inner}\rightarrow\text{outer}}$ should always be non-negative. Therefore, the second and third channels go through the ReLU activation to force the SDF difference maps $\Delta\hat{\phi}$ to be non-negative. The subtraction of the first and second channel is the predicted SDF of the arachnoid, while subtraction with the third channel is the SDF of the dura. These results are further compared with the pseudo ground truth obtained from NTGDM~\cite{Duan_SPIE_2022}. The $L1$ loss $\textit{l} = \frac{1}{N} \sum_{i} |x_{i} - y_{i}|$ was applied to train the network.

\noindent\textbf{Surface Reconstruction}~~~~After achieving the SDF of the dura and arachnoid, we adopt a connectivity consistent marching cubes~\cite{marching_cubes}  to obtain a triangular mesh representation of each meningeal surface. Marching cubes is an iterative surface reconstruction algorithm that repeatedly tests the sign of the 8 vertices the uniform grids. If all vertices on the grid are positive or negative, the entire grid is outside or inside the surface. The algorithm will construct the triangle elements within the cubes that have different signs of the SDF. In the connectivity-consistent marching cubes, the contour location is computed by linear interpolation of the SDF with the connectivity considered for ambiguous cases.

\section{Experiments and Results}
\noindent\textbf{Datasets}~~~~We conduct experiments on a normal aging cohort and a cohort of people with multiple sclerosis~(PwMS) cohort. The normal aging data comes from the Baltimore Longitudinal Study of Aging~(BLSA)~\cite{Shock-1984-BLSA}, and the PwMS cohort is collected on a Philips Achieva 3.0 T scanner. 28 male with age $78.2\pm5.4$ and 28 female with age $78.2\pm5.6$ subjects  were selected from BLSA, and 10 male with age $47.1\pm8.5$ and 22 female with age $49.7\pm10.8$ subjects from the PwMS cohort for ICV and SAS volumetric analysis.

Paired T1w and T2w images from both cohorts were run through a pre-processing pipeline: N4 bias correction~\cite{tustison2010n4itk}, rigid registration to an MNI-space, skull stripping, and white matter intensity normalization~\cite{reinhold2019evaluating}. The 2D acquired T2w images were super-resolved and anti-aliased using SMORE~\cite{zhao2020smore}. Pia surfaces were reconstructed via SLANT-CRUISE~\cite{Huo-2016-MACRUISE,Huo-2019-slant}

\noindent\textbf{Implementation Details}~~~~
The 2D U-Net was trained using Adam optimizer with a learning rate of $2 \times 10^{-4}$ for 60 epochs. The mini-batch size is 8. The initial number of channel in our U-Net is 64, with five downsamplingblocks The training
data~\cite{Mallika-2021-CT,Roy-2017-MONSTR}, containing 6 and
20 subjects for the arachnoid and dura surfaces, respectively,
were harmonized to the BLSA contrasts using
CALAMITI~\cite{ZUO_CALAMITI_2021}. Our approach completes the surface reconstruction in 4~min, approximately \textbf{3 times faster} than NTGDM. 



\begin{table*}[!tb]
  \caption{BLSA subjects and PwMS fixed-effect coefficients~($\beta$), standard error~(SE),
  and $p$-values~($p$) for the sex, baseline age, and follow-up interval.
  Statistically significant effects with $p$-values~$\leq 0.05$ are highlighted in bold.}
  \centering
  \begin{tabular}{l c ccc c ccc}
  \toprule

      &\multirow{2}{*}{\textbf{Subjects}}& \multicolumn{3}{c}{\textbf{ICV}} & \hspace*{3em}& \multicolumn{3}{c}{\textbf{SAS volume}} \\

      \cmidrule(l){3-5} \cmidrule(l){7-9}

    && $\boldsymbol{\beta}$ & \textbf{SE} & \textbf{\textit{p}-value} &&  $\boldsymbol{\beta}$ & \textbf{SE} & \textbf{\textit{p}-value}\\

   \midrule

\multirow{2}{*}{\textbf{Sex}} &\textbf{BLSA}& \textbf{144.60} & \textbf{28.91} & $\mathbf{6.87 \times 10^{-6}}$ && 9.54 & 11.51 &
0.41  \\
&\textbf{PwMS}& \textbf{260.039} & \textbf{36.371} & $\mathbf{7.207 \times 10^{-8}}$ && -1.453 & 18.463 & 0.938\\
\midrule
\multirow{2}{*}{\textbf{Baseline age}} &\textbf{BLSA}& -2.01 & 2.66 & 0.45 && \textbf{1.87} &
\textbf{0.87} & $\mathbf{3.66 \times 10 ^{-2}}$ \\
&\textbf{PwMS}& -0.566 & 1.901 & 0.768 && 0.611 &
0.880 & 0.493 \\
\midrule
\multirow{2}{*}{\textbf{Follow-up interval}} &\textbf{BLSA}& \textbf{-1.38} & \textbf{0.23}  & $\mathbf{1.35 \times 10^{-8}}$ &&
\textbf{2.42} & \textbf{0.41} & $\mathbf{2.82 \times 10 ^{-8}}$ \\
&\textbf{PwMS}& 1.13 & 0.711  & 0.114 &&
\textbf{1.02} & \textbf{0.465} & \textbf{0.03} \\
  \bottomrule
\end{tabular}
\label{tab:lme}
\end{table*}







\begin{table}[!tb]
\caption{Quantitative comparison between NTGDM and our proposed method. CG stands for curvature gradient. SD represents surface distance. Statistical analysis is conducted via paired Wilcoxon signed-rank test and $*$ marks the statistical significance with $p<0.05$.}
\centering
\resizebox{1\columnwidth}{!}{
\begin{tabular}{c c c c}
\toprule
\textbf{Method} & \textbf{Surface} & \textbf{Median CG}$\ \downarrow$ & \textbf{SD (mm)}$\ \downarrow$\\
\midrule
\multirow{2}{*}{\textbf{NTGDM}} & {\textbf{Arachnoid}} & $.086\pm.003 $ & $1.031\pm.270$ \\
& {\textbf{Dura}} & $.094\pm.004 $ & - \\
\multirow{2}{*}{\textbf{Ours}} & {\textbf{Arachnoid}} & \bm{$.021\pm.003^{*}$} & \bm{$.509\pm.249^{*}$} \\
& {\textbf{Dura}} & \bm{$.022\pm.003^{*}$} & - \\
\bottomrule
\end{tabular}
}
\label{table:comparison}
\end{table}

\begin{figure}[!tb]
\centering
  \includegraphics[width=\linewidth]{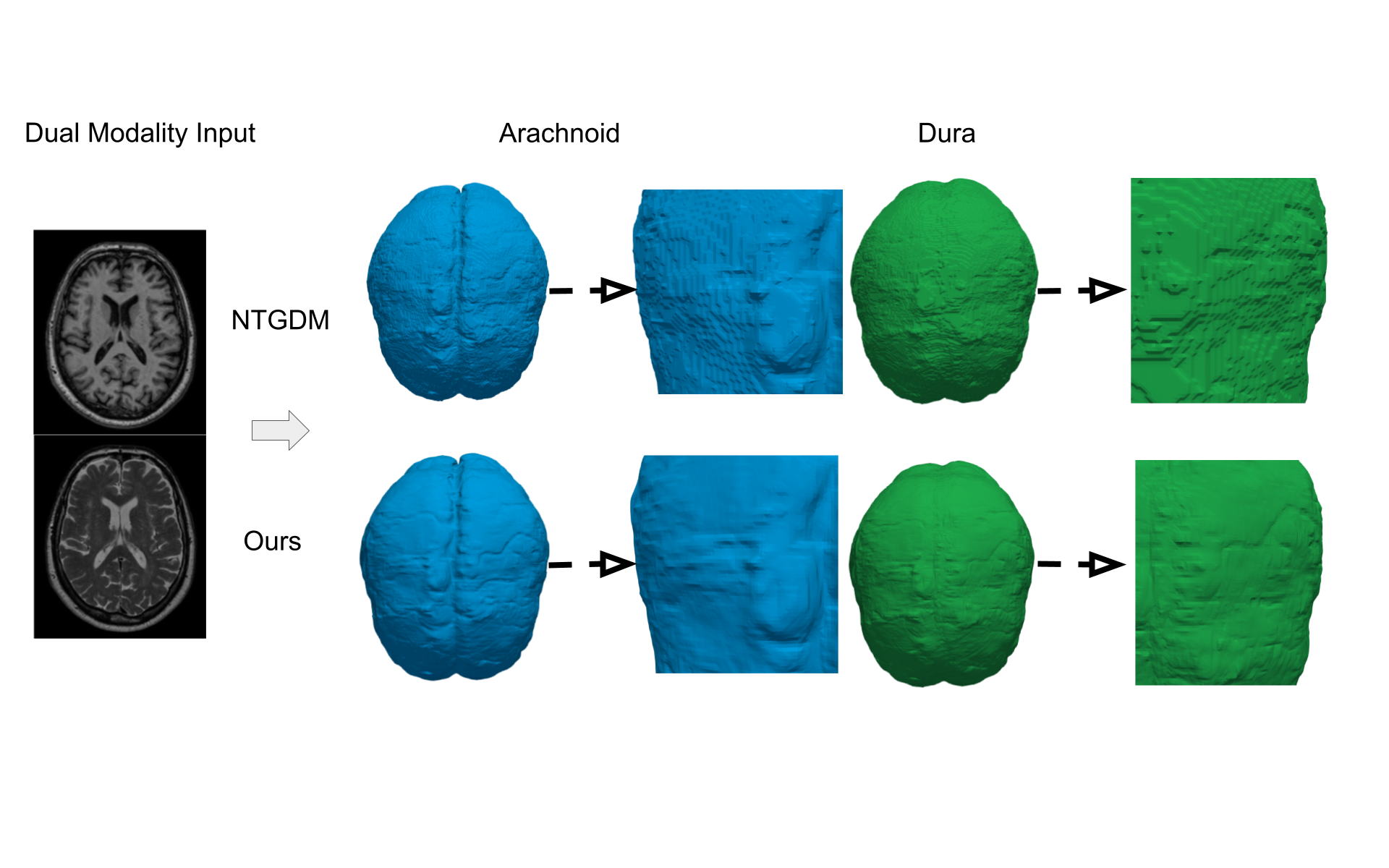}
  \caption{The reconstructed surface meshes of two algorithms are compared visually for the meningeal layers.
  }
  \label{fig:mesh result}
\end{figure}

\noindent\textbf{Evaluation metrics}~~~~To compare the smoothness of the reconstructed surfaces, we calculate the derivative of Gaussian curvature since the derivative of curvature is suggestive of sharp edges in 3D meshes~\cite{curvature_gradient}. Curvature gradient~(CG) is computed as,
\begin{equation*}
\frac{\partial K}{\partial x_{i}}=\sum_{j} J_{i j}^{-1} \frac{\partial K}{\partial r_{j}},
\quad\text{with}\quad
K\left(v_{i}\right)=2 \pi-n(v_{i})\theta(v_{i},f),
\end{equation*}
where $x_{i}$ is the global coordinate, $r_{j}$ is parametric coordinate and $J$ is the the Jacobian matrix, $n(v_{i})$ is the facet neighbours of the vertex $i$, $\theta(v_{i},f)$ is the angle of $f$ at vertex $i$.  Figure~\ref{fig:mesh result} demonstrates the reconstructed surface of one subject from the BLSA dataset between NTGDM~\cite{Duan_SPIE_2022} and our algorithm. One can observe that our reconstructed surface is clearly smoother than the NTGDM method.

To measure the accuracy of reconstructed surfaces, we manually annotated arachnoid surface points with sub-voxel accuracy on 5 BLSA subjects with 50 points on each subject, and calculate the distance between annotated points to the surface mesh, denoted as surfance distance~(SD). Quantitative comparisons between NTGDM and our method can be found in Table~\ref{table:comparison}. For both metrics, our method significantly outperforms the previous NTGDM-based method.

\begin{figure}[!t]
\centering
 \includegraphics[width = \linewidth]{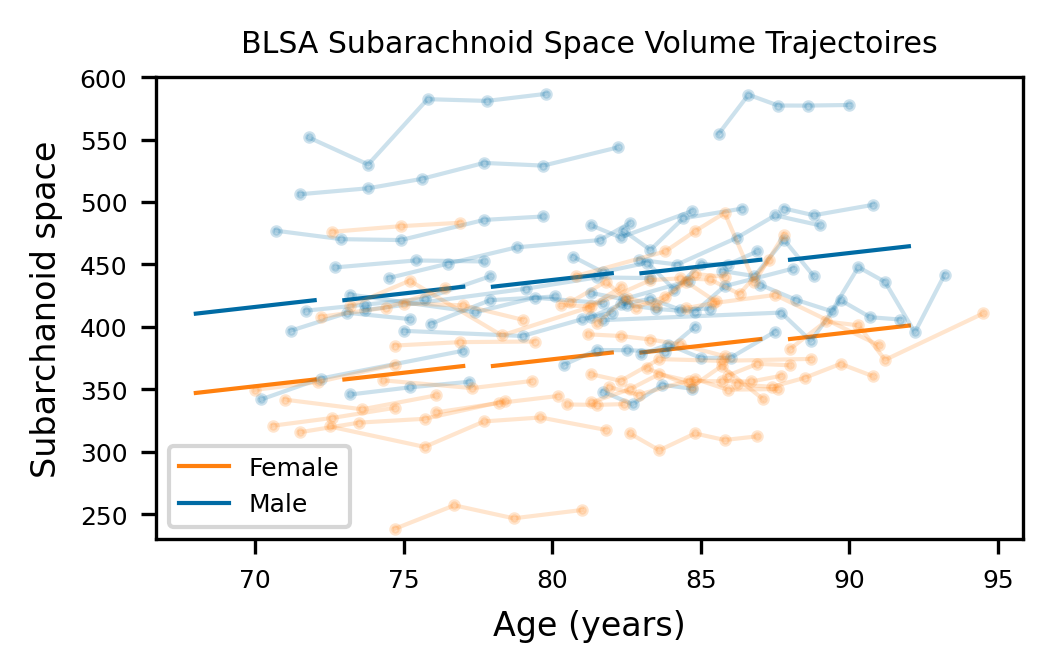}
 \includegraphics[width = \linewidth]{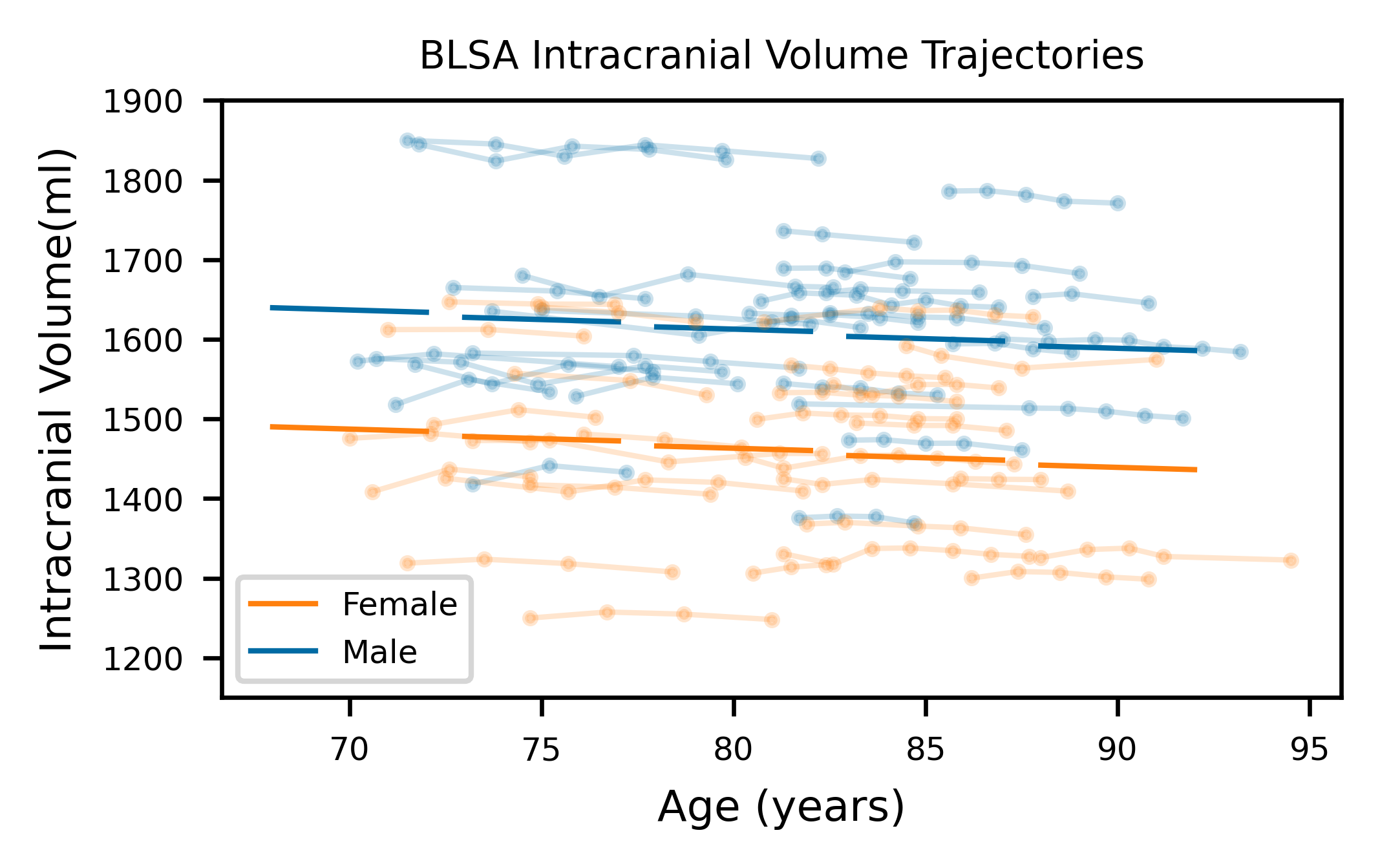}
  \caption{Individual volume of all BLSA subjects and visits.
  Each dot represents the volume measurement at a visit from an individual.
  Males and females are plotted in blue and orange, respectively. Each segment of trendline is calculated with a 5-year follow-up interval. SAS trendlines are adjusted with sex-specific average ICV values.}
  \label{fig:BLSA_Volume}
\end{figure}

\begin{figure}[!t]
\centering
  \includegraphics[width = \linewidth]{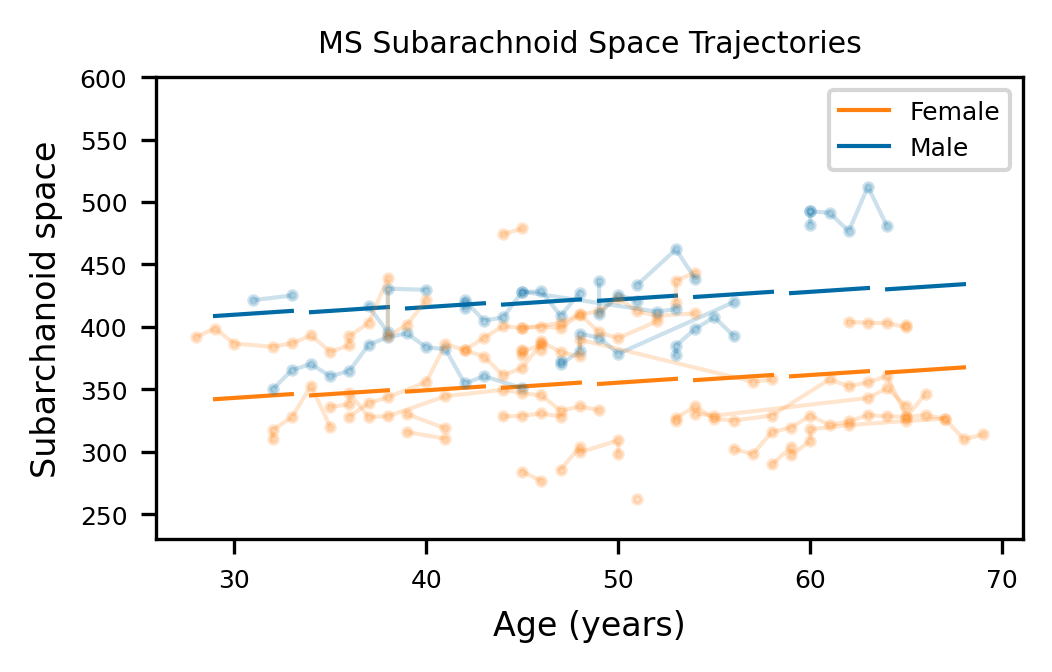}
  \includegraphics[width=\linewidth]{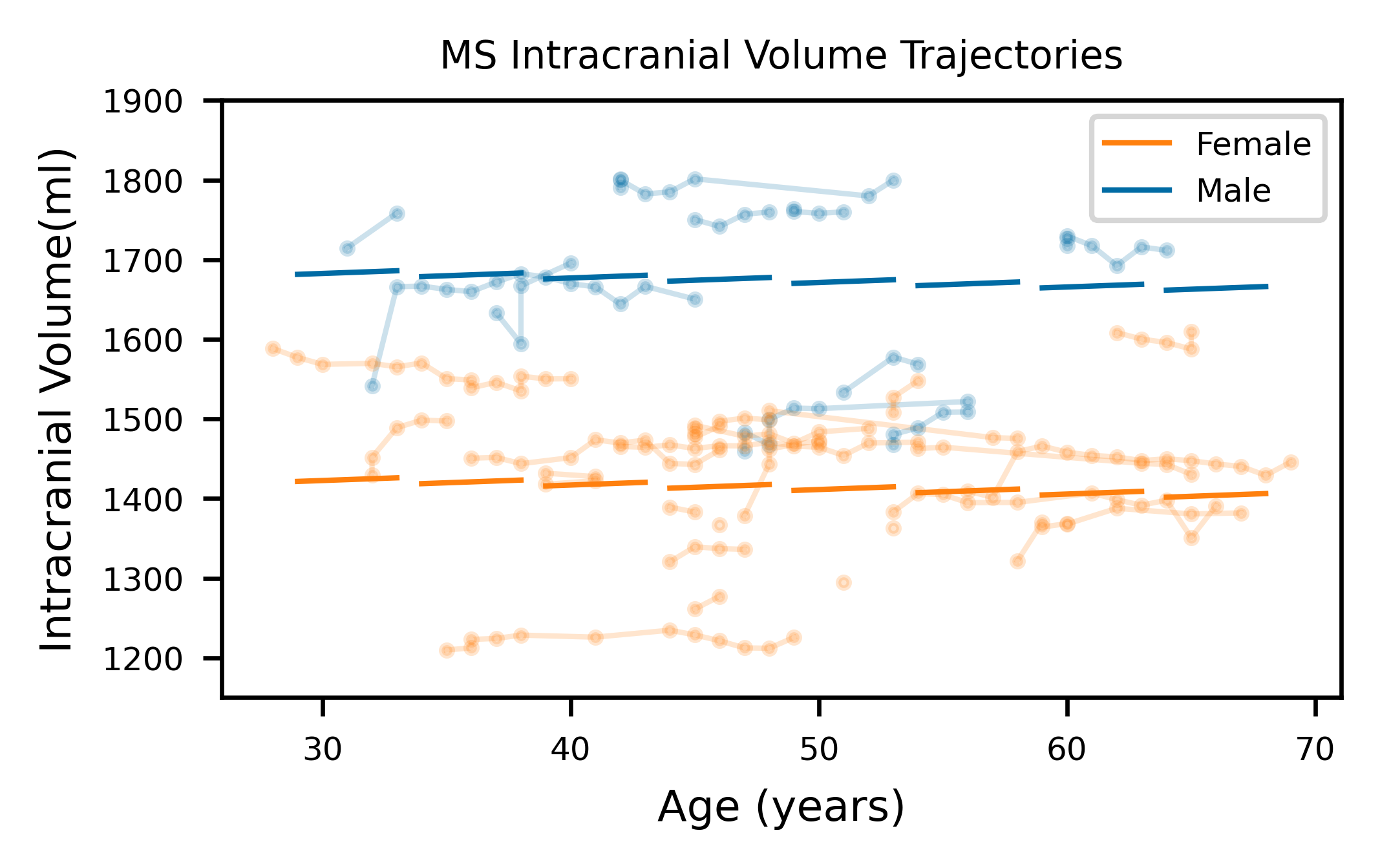}
  \caption{Individual volume of all visits for PwMS.
  Each dot represents the volume measurement at a visit from an individual.
  Males and females are plotted in blue and orange, respectively.}
  \label{fig:MS_Volume}
\end{figure}

\noindent\textbf{Longitudinal Analysis}~~~We chose 56 BLSA subjects with 212 total visits and 32 PwMS with 192 total visits for analyses. Linear mixed effects~(LME) model is used to study the relationship between age, sex, baseline, and longitudinal changes in SAS volumes and ICV for both BLSA subjects and PwMS \cite{Duan_SPIE_2022}. The fixed effects of SAS volume for both cohorts are intercept, baseline age, sex, follow-up interval, and ICV. The fixed effects for ICV are sex, baseline age, follow-up interval, and intercept. The random effects are intercept and follow-up interval for all analyses. The LME results are shown in Table~\ref{tab:lme}. The volume trajectories for all BLSA visits and PwMS visits are plotted in Fig.~\ref{fig:BLSA_Volume} and Fig.~\ref{fig:MS_Volume}, respectively. For BLSA subjects, we found that the ICV coefficients of both sex and follow-up interval significantly differs from 0 and SAS coefficents of both baseline age and sex significantly differs from 0. For PwMS, sex is a significant predictor for the ICV and follow-up interval is a significant predictor for SAS volume changes. 

%
%
%
%

\section{Discussion and Conclusions}
In this paper, we described a method to generate topology-preserving meningeal surfaces using CNNs. Compared to the previous state-of-the-art method, our proposed method is more computationally efficient and generates smoother as well as more precise meningeal surfaces with sub-voxel accuracy. We applied our proposed method on both healthy subjects and PwMS to study longitudinal changes of SAS volumes and ICVs. For our healthy control, our proposed method found statistical significance in both sex and follow-up intervals in ICV changes, meaning that head size differs significantly between females and males while decreasing over time. We also found statistical significance in both baseline age and follow-up interval for SAS volume changes, meaning that SAS increases both longitudinally and cross-sectionally during normal aging. For our MS subjects, we only found statistical significance in sex coefficients for ICV volumes, and follow-up interval in SAS volume, indicating that SAS volume increases longitudinally for PwMS. 

%

\section{Acknowledgments}
\label{sec:acknowledgments}

The authors would like to thank BLSA participants. This research was supported in part by the Intramural Research Program of the NIH, National Institute on Aging, and through NINDS grants R21~NS120286~(PI:~J.L.~Prince) and R01-NS105503~(PI: R.P.~Gullapalli).

\bibliographystyle{IEEEbib}
\bibliography{isbi2023_meninges}

\end{document}